\documentstyle[prl,aps,graphics,twocolumn]{revtex}
\input epsf
\begin{document}

% for the author list

  \newcommand{\bmit}{$^1$}
  \newcommand{\mainz}{$^2$}
  \newcommand{\csu}{$^3$}
  \newcommand{\tud}{$^4$}
  \newcommand{\uol}{$^5$}
  \newcommand{\uok}{$^6$}
  \newcommand{\fsu}{$^7$}
  \newcommand{\uog}{$^8$}

\draft

\title{High-precision Studies of the 
$^{\bf{3}}$He(e,e$^{\bf{\prime}}$p) Reaction at the Quasielastic Peak}

\author{
R.E.J.~Florizone,\bmit $ $ 
W.~Bertozzi,\bmit $ $
J.P.~Chen,\bmit$^,$\cite{PChen} $ $
D.~Dale,\bmit$^,$\uok,   $ $
S.~Gilad,\bmit $ $
A.J.~Sarty,\bmit$^,$\fsu, $ $
J.A.~Templon,\bmit$^,$\uog  $ $
S.P.~Van~Verst,\bmit$^,$\cite{SVerst} $ $
J.~Zhao,\bmit$^,$\cite{JZhao},  $ $
Z.-L.~Zhou,\bmit $ $
P.~Bartsch,\mainz   $ $
W.U.~Boeglin,\mainz$^,$\cite{PBoeglin} $ $
R.~B\"ohm,\mainz  $ $
M.O.~Distler,\mainz $ $
I.~Ewald,\mainz $ $
J.~Friedrich,\mainz $ $
J.M.~Friedrich,\mainz $ $
R.~Geiges,\mainz $ $
P.~Jennewein,\mainz $ $
M.~Kahrau,\mainz $ $
K.W.~Krygier,\mainz $ $
A.~Liesenfeld,\mainz $ $
H.~Merkel,\mainz $ $
K.~Merle,\mainz $ $
U.~M\"uller,\mainz $ $
R.~Neuhausen,\mainz $ $
E.A.J.M.~Offermann,\mainz$^,$\cite{POffermann} $ $
Th.~Pospischil,\mainz $ $
G.~Rosner,\mainz $ $
H.~Schmieden,\mainz $ $
A.~Wagner,\mainz $ $
Th.~Walcher,\mainz $ $
K.A.~Aniol,\csu  $ $
M.B.~Epstein,\csu  $ $
D.J.~Margaziotis,\csu $ $
M.~Kuss,\tud$^,$\cite{PChen} $ $
A.~Richter,\tud  $ $
G.~Schrieder,\tud $ $
K.~Bohinc,\uol  $ $
M.~Potokar,\uol  $ $
S.~\v{S}irca\uol $ $\\[3mm]
}

\address{
    {\bmit  \it  Laboratory for Nuclear Science, 
                 MIT, Cambridge, MA 02139, USA}\\
    {\mainz \it  Institut f\"ur Kernphysik, 
                 Universit\"at Mainz, D-55099 Mainz, Germany}\\
    {\csu   \it  Department of Physics and Astronomy, California State University
                 at Los Angeles, Los Angeles, CA 90032, USA}\\
    {\tud   \it  Institut f\"ur Kernphysik, Technische Universit\"at Darmstadt, 
                 D-64289 Darmstadt, Germany}\\
    {\uol   \it  Institute ``Jo\v{s}ef Stefan'', 
                 University of Ljubljana, SI-1001 Ljubljana, Slovenija}\\
    {\uok   \it  Department of Physics and Astronomy,
                 University of Kentucky, Lexington, KY 40506, USA}\\
    {\fsu   \it  Department of Physics,
                 Florida State University, Tallahassee, FL 32306, USA}\\
    {\uog   \it  Department of Physics and Astronomy,
                 University of Georgia, Athens, GA 30602, USA}\\
}

\date{\today}
\maketitle

\begin{abstract}
Precision studies of the reaction $^{3}$He(e,e$^\prime$p) 
using the three-spectrometer facility at the Mainz microtron MAMI
are presented. All data are for
quasielastic kinematics at $|\vec{q}\,| =685$ MeV/c.  
Absolute cross sections were measured at three electron kinematics.
For the measured missing momenta range from 10 to 165 MeV/c, 
no strength is observed for missing energies higher than 20 MeV.  
Distorted momentum distributions were extracted for the two-body breakup 
and the continuum.  
The longitudinal and transverse behavior was studied 
by measuring the cross section for three photon polarizations.  
The longitudinal and transverse nature of the cross sections 
is well described by a currently accepted and widely used prescription of the off-shell 
electron-nucleon cross-section.  The results are compared to modern 
three-body calculations and to previous data.\\
\end{abstract}
\pacs{PACS numbers: 21.45.+v,25.10.+s,25.30.Dh,25.30.Fj}
%PACS numbers: 21.45.+v, 25.10.+s, 25.30.Dh, 25.30.Fj

%\narrowtext
%\twocolumn
% body of paper here

The study of few-body nuclear systems has acquired new importance due to recent 
developments on both theoretical and experimental fronts.  
Several schemes have been developed to perform microscopic 
calculations which are based on the NN interaction rather than on a mean-field 
approach.  These include non-relativistic  
Faddeev-type calculations for three-body systems \cite{Golak,Kotlyar}
and Monte-Carlo variational calculations for three- and 
four-body systems \cite{Carlson}.  Fully-relativistic calculations 
are also being developed for three bodies \cite{Gross}. New experimental 
facilities with high-quality continuous-wave (cw) electron 
beams and high-resolution spectrometers provide the means to 
rigorously test these modern calculations. In particular, 
precision measurements of electromagnetic response functions, 
which are selectively sensitive to the various components of the nuclear 
currents, are possible.
% both for transition to bound final state configurations
%and to the continuum of unbound states.

The first results of a program to 
study $^{3,4}$He(e,e$^\prime$p) at a 
fixed 3-momentum transfer, $q = |\vec{q}\,| = 685$ MeV/c, 
and three energy transfers, $\omega$, 
corresponding to kinematics on top of the quasielastic peak, well above 
it (``dip'') and well below it are reported here.  
By varying the energy transfer, we hope to 
selectively enhance or suppress various effects contributing to the 
interaction.
In order to further understand these contributing effects, 
we studied the longitudinal and transverse components of 
the cross section by measuring it  at
 three electron scattering angles (virtual photon polarizations, $\epsilon$). 
The ongoing research program was carried out in the three-spectrometer 
facility\cite{Neuhausen} at the Mainz microtron MAMI by the A1 collaboration.  
The results reported here are 
from measurements performed on $^{3}$He in quasielastic kinematics 
($x_{\rm B}=1$ ; $\omega = 228$ MeV), where the dominant 
mechanism is expected to be the quasi-free knockout of a single proton.  
Further experimental details are described in Ref. \cite{Florizone}.

Few exclusive and semi-exclusive electron-scattering measurements
have been performed on $^3$He, and the existing data are
inconclusive.
The $^3$He(e,e$^\prime$p) reaction 
was measured at  Saclay \cite{jans} at 3-momentum transfers of 300
and 430 MeV/c.  Measurements covered the missing energy range 0--70 MeV
and missing momentum range 0 $\le p_m \le$ 300 MeV/c.  
Momentum distributions were
extracted for the two-body breakup channel, 
$^3$He(e,e$^\prime$p)$^2$H.  
Another $^3$He(e,e$^\prime$p) measurement 
at Saclay \cite{marchand} in dip kinematics suggested the existence
of two-body (short-range) correlations.  
The latter data exhibit some
agreement with calculations by Laget \cite{laget}.  Both experiments were
performed in perpendicular kinematics.  The longitudinal and transverse
response functions were measured for $^3$He(e,e$^\prime$p)$^2$H in 
quasielastic kinematics at 
Saclay \cite{ducret}
over the $q$ range of 350--700 MeV/c.  At the lower $q$, the longitudinal 
response 
is quenched by about 30\%
relative to the transverse, while at $q > 500$ MeV/c, the experimental
spectral functions are consistent with PWIA predictions.  In another L/T
measurement in the dip region \cite{legoff} at missing momentum 
260 MeV/c, the
longitudinal spectral function was observed to be strongly quenched for
both the two-body breakup and the continuum channels,
in agreement with calculations by Laget which include
meson-exchange currents and final-state interactions (FSI).  A few 
experiments are planned at the high momentum-transfer range now available at 
TJNAF.  In particular, a measurement of cross sections and response functions 
at high $q$ and for high $p_m$ \cite{e89-044} is scheduled for late 1999.

In this Letter, we report on measurements of the 
 $^3$He(e,e$^\prime$p) reaction 
%performed with the three-spectrometer setup at MAMI.
%\cite{Neuhausen} in the A1 Hall at the Mainz  Microtron 
%Facility in Mainz, Germany.
%  The measurements were performed
at a central 3-momentum transfer $q=685$ MeV/c and
 central energy transfer $\omega=228$ MeV, corresponding
to the center of the quasielastic peak. 
Three incident beam energies
 ($E_b=$ 540, 675, and 855 MeV)
with electron scattering angles
($\theta_e=$ 103.85$^\circ$, 72.05$^\circ$, and 52.36$^\circ$)
were used.  This corresponds to three virtual
photon polarizations ($\epsilon =$ 0.214, 0.457, 0.648) respectively.
Protons with momenta ranging from 393 to 710 MeV/c 
were detected in parallel kinematics 
$(\vec{p_p} \parallel \vec{q}\,)$ to facilitate the study of the longitudinal and transverse components of the cross section. 
The incident cw beam current was 40 $\mu$A, and the beam
was rastered
on the target by $\pm 3.5$ mm in both the
horizontal and
vertical directions.
Scattered electrons
were detected in Spectrometer A ($\Delta \Omega \approx 21$ msr)
and time-coincident protons in Spectrometer B ($\Delta \Omega \approx 5.6$ msr).
Spectrometer C was used as a luminosity monitor by detecting
electrons at a fixed setting for each kinematics.
The solid angle subtended by the spectrometers 
for an extended target was extensively studied, and a detailed description can
be found in Ref. \cite{Florizone}.

The target \cite{target1,target2} consisted  of cold $^3$He gas
($T = 20-23$K and $P=5-10$ atm)
 encapsulated in a
 stainless steel sphere
8 cm in diameter and having 82 $\mu$m thick walls.
A fan circulated the target gas from the cell
through a heat exchanger to dispense the heat
deposited by the incident electron beam.
The target density was determined at each beam energy
by measuring elastic electron scattering at
a low beam current (5 $\mu$A) and comparing 
the result to published cross sections \cite{Ottermann}.
The target density (which varied with the 
incident electron beam current)  was then 
monitored and determined  continuously 
using the singles rate in Spectrometer C.
The systematic error on the overall normalization
is $\pm$5\% 
and is dominated by the 
uncertainties in the $^3$He elastic cross sections
and the monitoring of the target density
over time.

At each kinematic setting, the $^3$He(e,e$^\prime$p)
cross section was measured as a 
function of both missing momentum ($p_m$) and missing energy 
($E_m$).  The missing momentum range accessed was different for each kinematic setting.  Hence, at $\epsilon =$ (0.214, 0.457, 0.648) the $p_m$ range was (10-95, 10-125, 10-165 MeV/c) respectively.  Great care was given to the radiative corrections,
 which were performed 
by unfolding \cite{unfolding} the radiative tails in 
a 2-dimensional $(E_m,p_m)$ space.
The correction factor proposed by
Penner \cite{Penner} was
 used for internal
(Schwinger) radiation, and
the factor proposed by Friedrich \cite{Friedrich} for external
radiation, which included substantial
contributions from the target walls.  
Additional details about the radiative corrections 
can be found in Ref. \cite{Florizone}.

\begin{figure}[htb]\unitlength1cm
\begin{picture}({6},{11.0})
\put(-0.5,+0.5){\epsfxsize=9.0cm \epsfbox{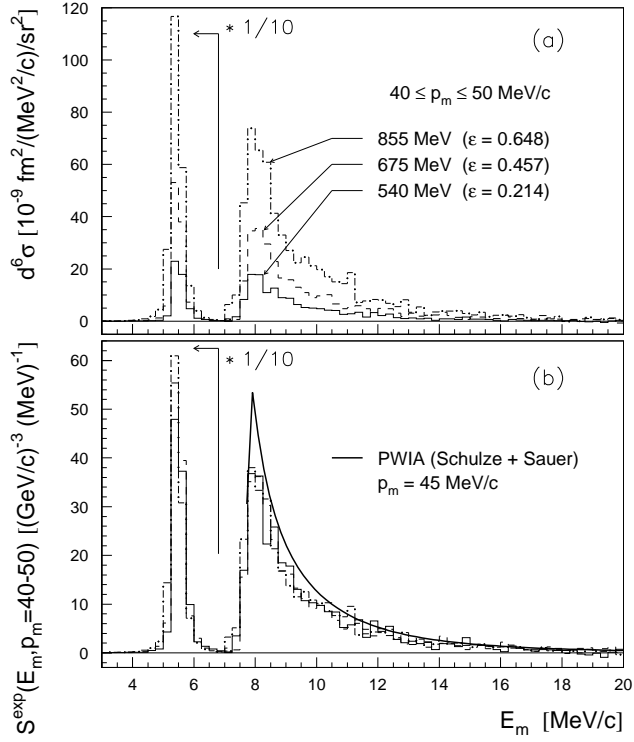}}
\end{picture}
\caption{
$^3$He(e,e$^\prime$p) cross sections (a) and spectral functions (b)
as functions of missing energy for a missing momentum bin 
$p_m$ = 40-50 MeV/c.
} \label{he3spec} \end{figure}

Fig. \ref{he3spec}(a) displays radiatively-corrected 
missing energy spectra for $^{3}$He(e,e$^\prime$p)  at the three different 
$\epsilon's$ and for an arbitrarily chosen missing-momentum  bin of 
40-50 MeV/c.  The high resolution evident in the figure enables accurate 
definition of the parameters ($E_m,p_m$) at which the cross 
sections were extracted.  The dominant features are the two-body breakup 
($^3$He $\rightarrow$ p + $^2$H)
peak at 5.49 MeV and the 
threshold for three-body breakup
($^3$He $\rightarrow$ p + p + n) at 7.72 MeV.
Higher missing energies correspond to 
the continuum of unbound states of the undetected
pn-pair.  The continuum includes the
excitation of the unbound singlet S-state
of the deuteron at $E_m = (7.72 + 0.55)$ MeV.
 There is no 
strength at $E_{m}>20$ MeV in the radiatively-corrected spectrum.  We note that 
there was no measured strength at $E_{m}>20$ MeV for the entire 
$p_{m}$ range for which these results are reported.  

The 2-dimensional 
experimental spectral function,
\begin{equation}
S^{\rm exp}(E_m,p_m) = \frac{1}{{p_p}^2 \cdot \sigma_{ep}^{CC1}} \cdot
\frac{d^6\sigma}{d\Omega_e d\Omega_p dE_e dp_p }\, , 
\end{equation}
 where $ \sigma_{ep}^{CC1} $ is the 
electron-nucleon off-shell cross section of de Forest \cite{deForest}
has been determined. 
Fig. \ref{he3spec}(b) displays the measured spectral functions
for the three virtual photon polarizations and for an arbitrary missing
momentum bin $40\le p_m \le 50$ MeV/c.
The extracted $^3$He(e,e$^\prime$p) 
spectral functions are found to be independent
of $\epsilon$  within 
statistical and systematic uncertainties over the entire
region $0 \leq E_m \leq 20$ MeV.
Also displayed in the figure 
is a theoretical spectral function from Schulze and Sauer \cite{Schulze} 
for $p_{m}=45$ MeV/c.  The theoretical spectral function describes 
the shape of the experimental one well, 
but the magnitude is approximately 20\% larger.  
Note that the experimental values were affected by FSI 
which account for at least part of this difference.

\begin{figure}[htb]\unitlength1cm
\begin{picture}({6},{16.0})
\put(-0.5,+7.4){\epsfxsize=9.0cm \epsfbox{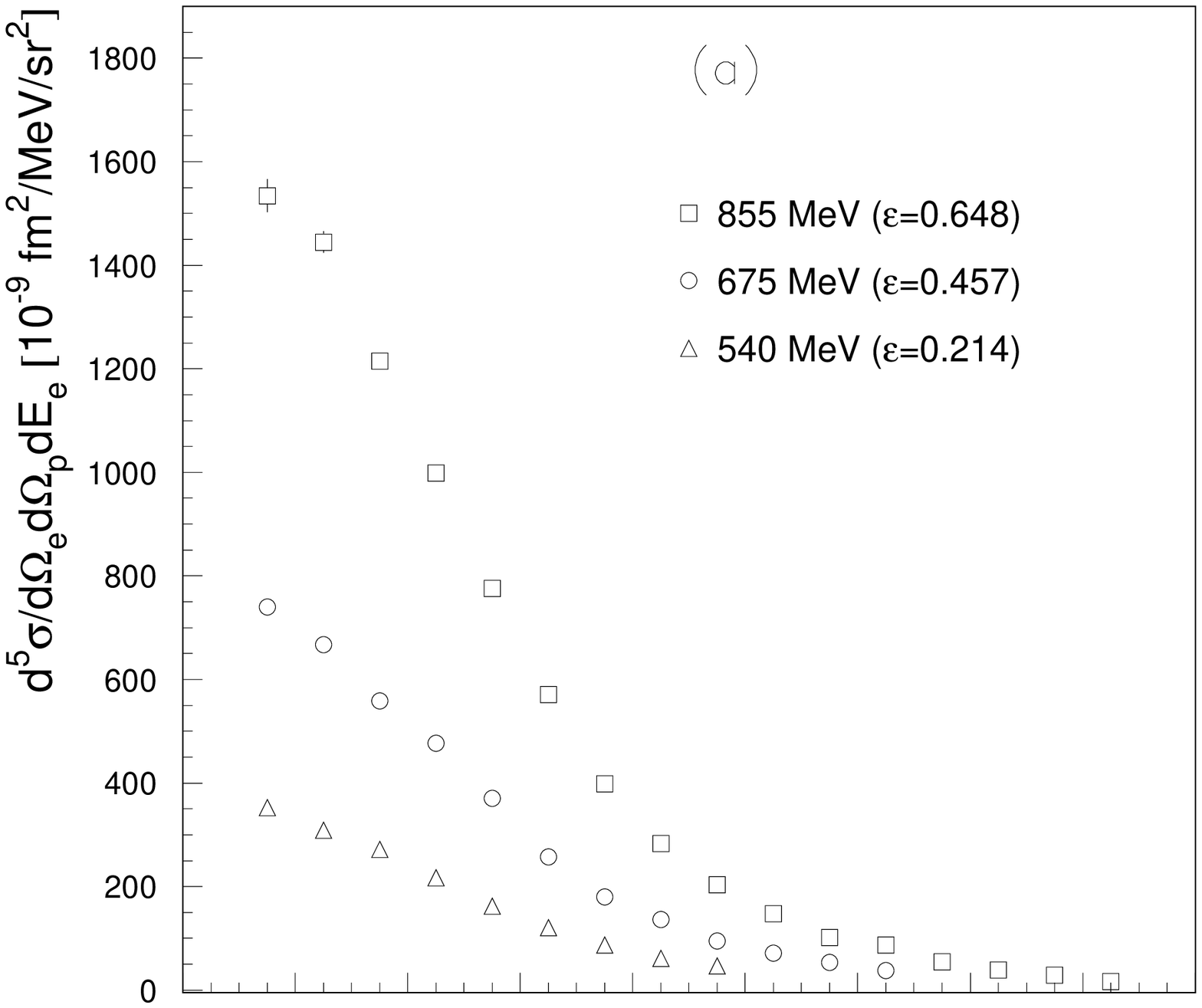}}
\put(-0.5,+0.5){\epsfxsize=9.0cm \epsfbox{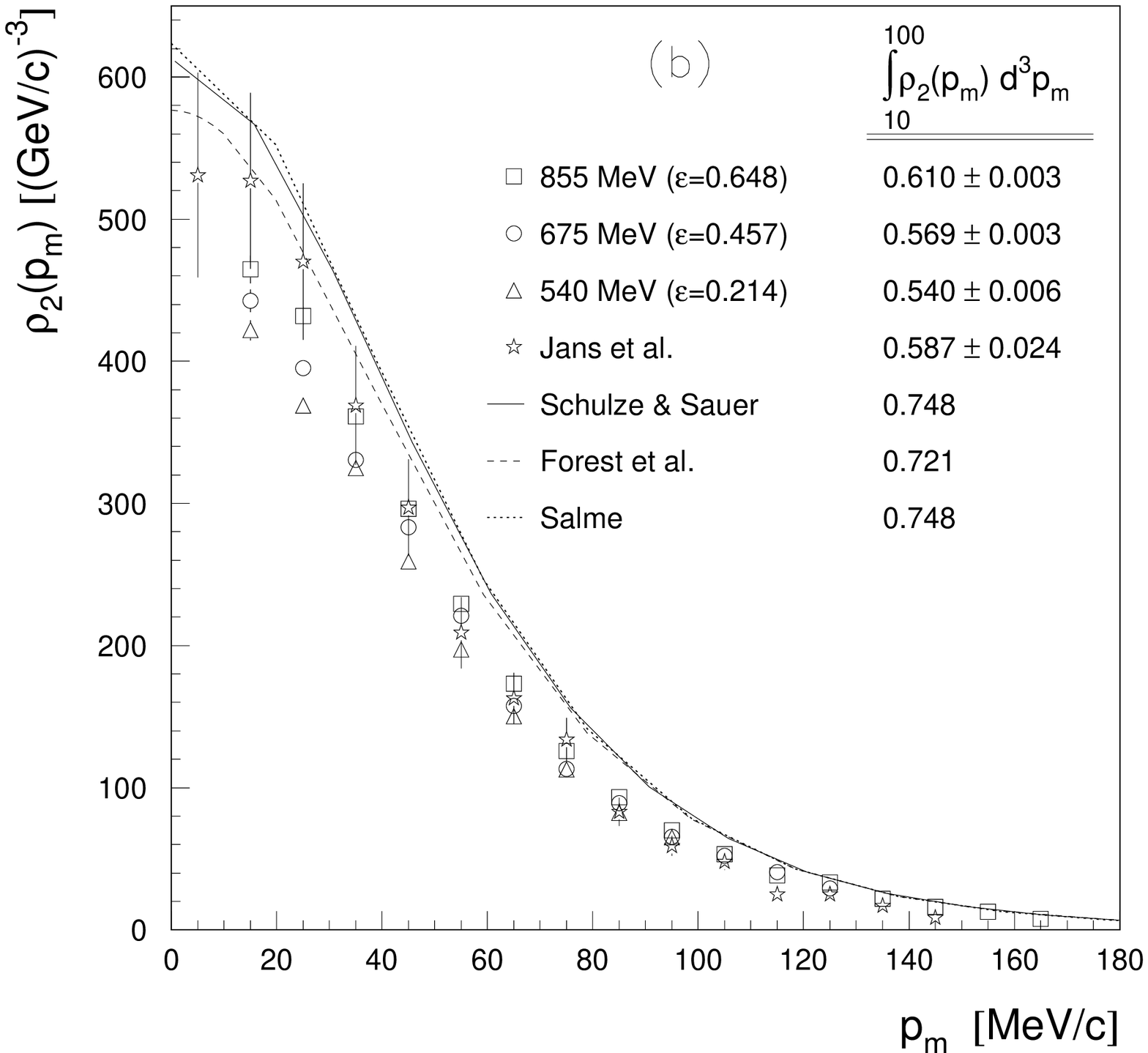}}
\end{picture}
\caption{
(a): $^3$He(e,e$^\prime$p)$^2$H cross sections
as a function of missing momentum. (b): extracted 
momentum distribution 
for the two-body breakup process
are compared to various calculations.
All uncertainties are statistical only.
}
\label{2body}
\end{figure}

For the two-body breakup channel, $^3$He(e,e$^\prime$p)$^2$H,
 the measured cross section as a function of $p_{m}$ for 
the three values of $\epsilon$ is shown in Fig. \ref{2body}(a).  
A strong dependence on $\epsilon$ is observed.  

To gain insight into the $\epsilon$-dependence
of the cross-section, the measured $^3$He 
proton momentum distributions  $\rho_2 (p_m)$, obtained by 
integrating $S^{\rm exp}(E_m,p_m)$ over the two-body breakup peak, 
are plotted in Fig. \ref{2body} (b).
The momentum distribution shows very little dependence on the virtual photon
polarization indicating that the 
longitudinal/transverse behavior of the two-body breakup cross
section is explained well by  $\sigma_{ep}^{CC1}$.
The data also overlap well with published results \cite{jans} 
obtained in non-parallel kinematics and at $q=430$ and 300 MeV/c, supporting
the (PWIA) hypothesis that S$^{\rm exp}$($E_m,p_m$)
can be factorized in the cross section.   

The remaining $\epsilon$ 
dependence of the cross sections 
is evaluated by comparing the integrals  
\begin{equation}
N(\epsilon)=4\pi \int_{10}^{100}\rho_{2}(p_{m})\cdot p_{m}^{2}dp_{m}
\end{equation}
in Fig. \ref{2body} (b). 
The uncertainties quoted for $N(\epsilon)$  are statistical only.
The values of $N(\epsilon)$ vary by about 10\% 
for $\epsilon=0.214-0.648$.  
We note that, to the extent this difference may be significant, 
the cross sections are slightly more longitudinal than 
those of $\sigma_{ep}^{CC1}$.  Also shown in the figure are three 
calculations by Schulze and Sauer \cite{Schulze}, 
Salm\`e {\it et al.} \cite{Kievsky}, 
and Forest {\it et al.} \cite{Forest96}.  
The Schulze and Sauer momentum distribution was 
calculated using the Paris potential, and is very similar to that of Salm\`e.  
The calculations by Forest {\it et al.} use the Argonne v18 NN-potential 
and the Urbana IX three-nucleon-interaction potential 
together with variational Monte-Carlo techniques.  
Note that the theoretically-extracted momentum distributions 
do not take into account final-state interactions which do affect the 
measured distributions.  
Hence, the differences of about 20-25\% 
between the measured and calculated integrals, N, in Fig. \ref{2body} (b)
contain (but are not necessarily restricted to) the effects of FSI.

\begin{figure}[htb]\unitlength1cm
\begin{picture}({6},{9.0})
\put(-0.5,+0.5){\epsfxsize=9.0cm \epsfbox{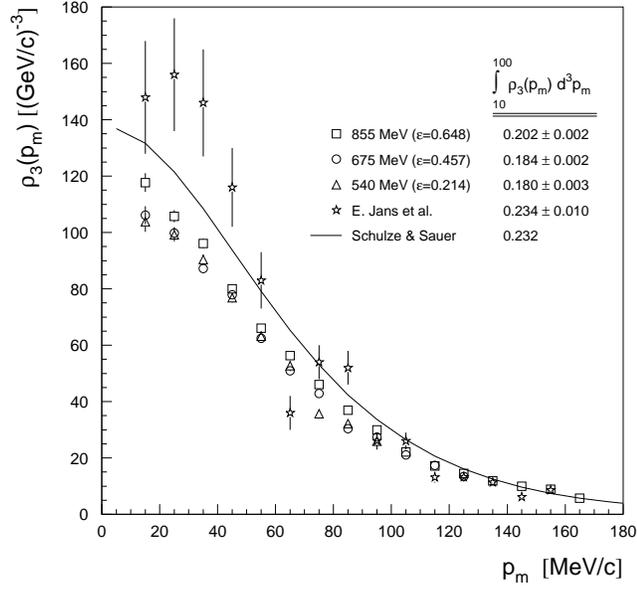}}
\end{picture}
\caption{
Momentum distribution for the 
$^3$He(e,e$^\prime$p)np continuum integrated over the range $E_m$ = 7-20 MeV.
All uncertainties are statistical only.
}
\label{3body}
\end{figure}

A similar analysis can be performed at higher excitation energies.  
The experimentally-extracted 
momentum distributions for the continuum, 
$\rho_{3} (p_m)$, integrated over the range $E_m$ = 7-20 MeV,  
are presented in Fig. \ref{3body}. 
They  display about a 10\% dependence on $\epsilon$,
similar to the two-body breakup channel.
Also shown in the figure are  older 
data from Saclay \cite{jans} and a calculated momentum distribution from 
Schulze and Sauer \cite{Schulze}.  We note that at $p_{m}<50$ MeV/c, our
data set is well below that of Jans {\it et al}.  
The theoretical values from 
Schulze and Sauer are again approximately 20\% 
larger than the data which are subject to the effects of FSI.

We have performed precise measurements of the 
$^3$He(e,e$^\prime$p) reaction in quasielastic kinematics.  
The data span the missing momentum range
up to 95-165 MeV/c, and missing energy range up to 80 MeV.  After radiative
corrections, there is no observed strength at excitations higher than 20 MeV.
The dependence on the
virtual-photon polarization of the observed cross section (and hence the 
longitudinal/transverse behavior) over the
entire excitation range 
is almost entirely due to the off-shell e-N cross section, and
is described well by $\sigma_{ep}^{CC1}$.  We conjecture that meson exchange and isobar currents, which are predominantly transverse, cannot be very important because the experimental results scale as $\sigma_{ep}^{CC1}$.  We note that in a similar region of momentum transfer ($q = 500$ MeV/c and $q = 1050$ MeV/c), the inclusive $^{3}$He(e,e$^\prime$) cross section, which integrates mainly over non-parallel kinematics, exhibits similar behavior \cite{inclusive1,inclusive2,inclusive3}.  Theoretical calculations of the 
spectral function and momentum distributions reproduce well the shape of 
the measured quantities, but are 10-20\%
larger in magnitude.  This difference can be attributed, at least in part, 
to the effects of FSI.
A theoretical study which includes an exact treatment of FSI 
is now in progress \cite{Ziemer}.

We would like to thank the MAMI staff for 
providing excellent support for this experiment.
This Work was supported by the Deutsche Forschungsgemeinschaft (SFB 201),
by the Bundesministerium f\"ur Forschung und Technologie
and by the U.S. Department of Energy and the National Science Foundation.

\end{document}